\documentclass[a4paper,12pt, epsfig]{article}
\pdfoutput=1
\usepackage{epsfig}
\usepackage{epstopdf}
\usepackage{graphicx}
\usepackage{ifthen}
\usepackage{feynmp}
\usepackage{ulem}

\usepackage{amsmath}
\usepackage[psamsfonts]{amssymb}
\usepackage{euscript}

\usepackage{latexsym}
\usepackage[arrow,matrix,curve]{xy}

\newskip\humongous \humongous=0pt plus 1000pt minus 1000pt

\newif\ifdtup

\def\,{\hspace{-.1cm}}
\def\hsp{,\hspace{.7cm}}

\def\fc#1#2 {\frac{n}{q}#1\frac{n}{q}#2}

\renewcommand{\cos}{\textrm{cos}}

\renewcommand{\theequation}{\arabic{section}.\arabic{equation}}
\renewcommand{\(}{\begin{equation}}
\renewcommand{\)}{end{equation} \vspace{-.05in}\linebreak}

\newcounter{saveeqn}
\newcounter{savealpheqn}

\newcommand{\alpheqn}{\setcounter{saveeqn}{\value{equation}}%
  \stepcounter{saveeqn}\setcounter{equation}{0}%
  \renewcommand{\theequation}{\mbox{\arabic{section}.\arabic{saveeqn}
\alph{equation}}}
  \renewcommand{\)}{\end{equation}}}
\def\part#1{\frac{\partial}{\partial{#1}}}%
\def\group#1{\refstepcounter{equation}\setcounter{saveeqn}
 {\value{equation}}%
  \label{#1}\setcounter{equation}{0}%
\renewcommand{\theequation}{\mbox{\arabic{section}.\arabic{saveeqn}
\alph{equation}}}
  \renewcommand{\)}{\end{equation}}}
\newcommand{\reseteqn}{\setcounter{equation}{\value{saveeqn}}%
  \renewcommand{\theequation}{\arabic{section}.\arabic{equation}}%
  \renewcommand{\)}{\end{equation}}}

\newcommand{\aalpheqn}{\setcounter{saveeqn}{\value{equation}}%
  \stepcounter{saveeqn}\setcounter{equation}{0}%
  \renewcommand{\theequation}{\mbox{
        \Alph{subsection}.\arabic{saveeqn}\alph{equation}}}
   \renewcommand{\)}{\end{equation}}}
\newcommand{\areseteqn}{\setcounter{equation}{\value{saveeqn}}%
  \renewcommand{\theequation}{\Alph{subsection}.\arabic{equation}}%
  \renewcommand{\)}{\end{equation}}}

\renewcommand{\thefootnote}{\alph{footnote}}
\renewcommand{\(}{\begin{equation}}
\renewcommand{\)}{\end{equation}}
\newcommand{\ba}{\begin{eqnarray}}
\newcommand{\ea}{\end{eqnarray}}
\newcommand{\cbp}{\mathop{\vtop{\ialign{##\crcr
   $\hfil\displaystyle{}\hfil$\crcr\noalign{\kern-13pt\nointerlineskip}
   \BIG{)}\hskip0pt\crcr\noalign{\kern3pt}}}}}
\newcommand{\pa}{\mathop{\vtop{\ialign{##\crcr
    
$\hfil\displaystyle{\oplus}\hfil$\crcr\noalign{\kern+1pt\nointerlineskip 
}
   \hspace{.08in}$^{\alpha=0}$\hskip6pt\crcr\noalign{\kern3pt}}}}}
\renewcommand{\hsp}{,\hspace{.3in}}
\catcode`\@=11
\def\vereq#1#2{\lower3pt\vbox{\baselineskip1.5pt \lineskip1.5pt
\ialign{$\m@th#1\hfill##\hfil$\crcr#2\crcr\sim\crcr}}}
\catcode`\@=12

\renewcommand{\(}{\begin{equation}}
\renewcommand{\)}{\end{equation}}

\newcommand{\E}{{\mathcal{E}}}
\newcommand{\beas}{\begin{eqnarray*}}
\newcommand{\eeas}{\end{eqnarray*}}

\newcommand{\bquo}{\begin{quote}}
\newcommand{\enqu}{\end{quote}}

\newcommand{\beq}{\begin{equation}}
\newcommand{\eeq}{\end{equation}}
\newcommand{\bea}{\begin{eqnarray}}
\newcommand{\eea}{\end{eqnarray}}

\newskip\humongous \humongous=0pt plus 1000pt minus 1000pt

\newif\ifdtup

\catcode`\@=11

\@addtoreset{equation}{section}

\def\@normalsize{\@setsize\normalsize{15pt}\xiipt\@xiipt
\abovedisplayskip 14pt plus3pt minus3pt%
\belowdisplayskip \abovedisplayskip
\abovedisplayshortskip \z@ plus3pt%
\belowdisplayshortskip 7pt plus3.5pt minus0pt}

\def\small{\@setsize\small{13.6pt}\xipt\@xipt
\abovedisplayskip 13pt plus3pt minus3pt%
\belowdisplayskip \abovedisplayskip
\abovedisplayshortskip \z@ plus3pt%
\belowdisplayshortskip 7pt plus3.5pt minus0pt
\def\@listi{\parsep 4.5pt plus 2pt minus 1pt
      \itemsep \parsep
      \topsep 9pt plus 3pt minus 3pt}}

\relax

\catcode`@=12

\catcode`\@=11

\def\section{\@startsection{section}{1}{\z@}{3.5ex plus 1ex minus  .2ex}{2.3ex plus .2ex}{\large\bf}}

\def\thesection{\arabic{section}}
\def\thesubsection{\arabic{section}.\arabic{subsection}}

\def\appendix{\setcounter{section}{0}
 \def\thesection{Appendix \Alph{section}}
 \def\thesubsection{\Alph{section}.\arabic{subsection}}
 \def\theequation{\Alph{section}.\arabic{equation}}}
\renewcommand{\theequation}{\arabic{section}.\arabic{equation}}

\begin{document}
% ======================================================================== 
\def\thefootnote{\fnsymbol{footnote}}
\def\thetitle{Measuring Entangled Neutrino States in a Toy Model QFT}
\def\autone{Hosam Mohammed}
\def\affa{Institute of Modern Physics, NanChangLu 509, Lanzhou 730000, China}
\def\affb{University of the Chinese Academy of Sciences, YuQuanLu 19A, Beijing 100049, China}
\def\affc{Physics department, Faculty of Science, Beni Suef University, Beni Suef, Egypt. 62514}

\begin{center}
{\large {\bf \thetitle}}

\bigskip

\bigskip

{\large \noindent  \autone{${}^{1,2}$}\footnote{hosam@impcas.ac.cn}, Jarah Evslin{${}^{1,3}$}\footnote{jarah@impcas.ac.cn} and Emilio Ciuffoli{${}^{1}$}\footnote{emilio@impcas.ac.cn}} %and \auttwo{${}^3$} \footnote{baiyang.zhang@wigner.mta.hu } }

\vskip.7cm

1) \affa\\
2) \affc\\
2) \affb\\
%3) \affc\\

\end{center}

\begin{abstract}
\noindent
Localized wave packet treatments of neutrino oscillations by various groups lead to mutually inconsistent predictions.   The neutrino wave packet description arises as an approximate substitute for the evolution of an entangled state which is not localized.    The disagreements arise from qualitative differences in the framework which are not specific to electroweak interactions, and so have analogues in simpler models.  Therefore in this note we introduce a toy model which allows one to explicitly test these predictions while consistently  keeping track of the entanglement of the neutrinos and the source particles.  Our study is robust as we use only the Schrodinger picture evolution equations on the entangled state, which are solved explicitly without recourse to the wave packet approximation.

\end{abstract}

% \vfill
% 
% \end{titlepage}
\setcounter{footnote}{0}
\renewcommand{\thefootnote}{\arabic{footnote}}

% \pacs{??}

%\ifthenelse{\equal{\pr}{1}}{
%\maketitle
%}{}

%\pr
%\ifthenelse{\equal{\pr}{1}}{yippy}

\section{Introduction}

In the standard, wave packet treatment of neutrino oscillations \cite{beuthe,giunti2012}, neutrino wave packets of each mass eigenstate are created simultaneously at the same position.  They then separate with time leading to decoherence.  On the other hand, in Ref.~\cite{mcgreevy}, the authors argue that the neutrinos are measured in a flavor eigenstate at a fixed moment in space and time, and so the mass eigenstates must have been emitted separately and then coelesced.  Clearly at most one of these discriptions is realized in Nature, but the neutrino wave packet description does not allow one to determine which.  Similarly in Ref.~\cite{dayadec} the Daya Bay collaboration analyzed their data using the oscillation probability found in Ref.~\cite{naumov1}, which they note is distinct from that of Ref.~\cite{beuthe}.  The difference, as is described in Ref.~\cite{naumov2}, arises from a difference in the assumptions regarding the kind of wave packet considered.  Again the wave packet model gives no guidance as to the initial conditions of the wave packets.

This situation arises because the wave packet is only an approximation to the true state of the quantum field theory.  As has been stressed in Ref.~\cite{cgl}, the true state is actually an entangled state of the neutrino with various particles involved in its production.  The wave packet approximation arises by truncating this entangled state to break the entanglement, however in Nature the entanglement is preserved\footnote{The standard argument, as reviewed in Ref.~\cite{giunti2012}, is that entanglement with the environment is similar to measuring the particles involved in the neutrino production, and so their wave functions are fixed and factored out.  Alternately, in \cite{revival} the effects of environmental interactions are incorporated by simply introducing a cut off in the production time.  However in a more complete treatment the entanglement with the environment may be considered explicitly.  Our model is designed to incorporate environmental interactions, as was shown in Ref.~\cite{noi1}.}.  Therefore to resolve the above issues, we feel that a treatment is required which uses the full entangled states, with no arbitrary truncation.  On the other hand, these issues depend only on the nature of wave packets and not the specifics of the electroweak interaction of interest.  Therefore we will consider a simplified model, in which the electroweak interactions are replaced by bosonic interactions in 1+1 dimensions.  In this simple model, we can explicitly solve the evolution equations for the full entangled state and so robustly determine its evolution and the result of measurements.

The program  was initiated in Ref.~\cite{noi1} where we introduced a model of neutrino production.  This alone is not sufficient to compare even qualitatively with neutrino oscillation phenomenology, as detection was not incorporated into the model.  In the current note we complete the model of \cite{noi1} by including neutrino detection.  As this model is not the electroweak model, quantitative predictions for neutrino oscillations will not be possible.  However, as is explained in Ref.~\cite{naumov2}, even the most qualitative features of neutrino oscillations are disputed in the wave packet formalism.  Our model is sufficient to distinguish between these various predictions and so to determine which wave packet treatment, if any, best approximates the true entangled evolution.  This will be done in the next paper in our series, the current paper introduces and tests the model itself.

We define the model in Sec.~\ref{modsez}.  Then in Sec.~\ref{ansez} we use the Hamiltonian evolution to derive a closed form expression for the probability of detecting a neutrino as a function of position.   In Sec.~\ref{numsez} we fix the parameters of the detector so as to maximize its detection efficiency and, for a fixed set of parameters, we numerically evaluate the appearance and disappearance channel oscillation probabilities as a function of the baseline $x$ between the neutrino source and detector.  Finally in Sec.~\ref{semisez} we develop two semianalytic approximations which roughly reproduce our results.  Sec.~\ref{litsez} describes other approaches to neutrino decoherence in the literature.

\section{The Model} \label{modsez}

\begin{table}
\begin{tabular}{|l|l|}
\hline
Field&Description\\
\hline
$\phi_{S L}$&The light source field \\
$\phi_{S H}$&The heavy source field \\
$\phi_{D L}$&The light dector field \\
$\phi_{D H}$&The heavy detector field \\
$\psi_{1}$&The light neutrino field \\
$\psi_{2}$&The heavy neutrino field \\
\hline
%\end{tabular}
%\begin{tabular}{|l|l|}
\hline
Symbol&Description\\
\hline
$M_{\alpha I}$&The mass of $\phi_{\alpha I}$ \\
$m_{i}$&The mass of $\psi_i$ \\
$E_{\alpha I}$&The on-shell energy of $\phi_{\alpha I}$\\
$e_{i}$&The on-shell energy of $\psi_i$\\
$\E_0$&Total on-shell energy before neutrino emission\\
$\E_{1i}$&Total on-shell energy between emission and absorption of $\psi_i$ (Process A)\\
$\tilde{\E}_{1i}$&Total on-shell energy between emission and absorption of $\psi_i$ (Process B)\\
$\E_{2}$&Total on-shell energy after neutrino absorption\\
$F$&A function of the $\E$s which oscillates rapidly if far off-shell\\
$\mathcal{A}(k,l)$&Amplitude for a final state with source (detector) momentum $l$ ($k$)\\
$\mathcal{P}(k,l)$&Probability density for a final state with source (detector) momentum $l$ ($k$)\\
$\mathcal{P}(k)$&Probability density for a final state with detector momentum $k$\\
\hline

\end{tabular}
\caption{Summary of Notation}
\end{table}

We consider a model in (1+1)-dimensions consisting of six real scalar fields.  At the source there is a heavy and light field $\phi_{SH}$ and $\phi_{SL}$ and at the detector a heavy and light field $\phi_{DH}$ and $\phi_{DL}$.  These have masses $M_{SH},\ M_{SL},\ M_{DH}$ and $M_{DL}$ respectively.  There are also two scalar fields with mass eigenstates $\psi_1$ and $\psi_2$ which will play the role of the neutrino.  Their masses will be denoted $m_1$ and $m_2$.   We will refer to these scalars as neutrinos.

The evolution of these fields is described by the standard free massive scalar Hamiltonian $H_0$ plus an interaction term
\beq
H_I=\int dx :\mathcal{H}_I(x):\hsp
\mathcal{H}_I(x)=\sum_{\alpha=\{S,D\}}\phi_{\alpha H}(x)\phi_{\alpha L}(x)\left(\psi_1(x)+\psi_2(x)\right). \label{hi}
\eeq
Here the sum $\psi_+=\psi_1+\psi_2$ plays the role of a flavor eigenstate and :: denotes the standard normal ordering of the creation and annihilation operators defined below.   We remind the interested reader that in 1+1 dimensions, normal-ordering of such theories renders them finite and so no renormalization is necessary.   The state $\psi_-=\psi_1-\psi_2$ plays the role of the other flavor eigenstate.  Thus the decay of the heavy source particle creates $\psi_+$ which later oscillates to $\psi_-$ and back.  We will consider the disappearance channel in which $\psi_+$ is measured and more briefly the appearance channel, in which $\psi_-$ is measured.  

Note that our toy model neutrinos are created in two body decays, such as the decay $\pi^+\rightarrow \mu^+ +\nu_\mu$.  However the kinematics of the decay will not be essential, and we are in fact motivated by reactor neutrinos.  In this case $\phi_{SH}$ represents a nucleus which $\beta$ decays to $\psi_{SL}$ plus an electron antineutrino and an electron.  The electron is not present in our model.  The electron antineutrino corresponds to $\psi_+$.   Imagine that there were only two flavors of neutrino.  Then $\psi_1$ and $\psi_2$ would be the two mass eigenstates.  The other flavor eigenstate, for example the muon antineutrino, would be $\psi_-$.    Similarly we identify $\phi_{DL}$ with a free proton which detects electron antineutrinos via inverse $\beta$ decay, producing $\phi_{DH}$ which represents the neutron.  The positron is not included in our model.

One may object that $\psi_-$ cannot be produced and so our mass matrix appears to be nonunitary. In the real world a $\beta$ decay may produce a muon antineutrino but it would be hopelessly off-shell because a muon must be produced simultaneously.  There is no muon in our model, but if desired we could mock up this effect in our model by including another coupling of $\phi_{SH}$ to $\psi_-$ and also a new very heavy $\phi_{SVH}$ which plays the role of the muon, and so guarantees that any $\psi_-$ produced by $\phi_{SH}$ decay will be hopelessly off-shell.  The same may be done with the detector particles.  In this way, our model may be given an ultraviolet completion in terms of a unitary mass matrix.  However we will not be interested in an ultraviolet completion of our model in this paper.

The canonical scalar fields will be decomposed in the Schrodinger picture as
\bea
\phi_{\alpha I}(x)&=&\int\frac{dp}{2\pi}\frac{1}{\sqrt{2E_{\alpha I}(p)}}\left(A_{\alpha I,-p}+A^\dagger_{\alpha I,p}\right)e^{-ipx}\hsp E_{\alpha I}(p)=\sqrt{M_{\alpha I}^2+p^2}\nonumber\\
\Pi_{\alpha I}(x)&=&-i\int\frac{dp}{2\pi}{\sqrt{\frac{E_{\alpha I}(p)}{2}}}\left(A_{\alpha I,-p}-A^\dagger_{\alpha I,p}\right)e^{-ipx}\nonumber\\
\psi_i(x)&=&\int\frac{dp}{2\pi}\frac{1}{\sqrt{2 e_i(p)}}\left(a_{i,-p}+a^\dagger_{i,p}\right)e^{-ipx}\hsp e_i(p)=\sqrt{m_i^2+p^2}\nonumber\\
\pi_i(x)&=&-i\int\frac{dp}{2\pi}{\sqrt{\frac{e_i(p)}{2}}}\left(a_{i,-p}-a^\dagger_{i,p}\right)e^{-ipx}.
\eea
We remind the reader that in the Schrodinger picture even interacting fields admit such a decomposition with $A$ and $A^\dagger$ and also $a$ and $a^\dagger$ satisfying the Heisenberg algebra.  Here and throughout the paper Greek indices run over $S$ and $D$, capital Roman indices over $H$ and $L$ and lower case Roman indices over $1$ and $2$.

We will be interested in states that have precisely one source particle, one detector particle and zero or one neutrinos.  These will be denoted respectively by
\beq
|I,p_2;J,p_1\rangle=A^\dagger_{DI,p_2}A^\dagger_{SJ,p_1}|\Omega\rangle\hsp
|I,p_2;J,p_1;i,q\rangle=a^\dagger_{i,q}|I,p_2;J,p_1\rangle
\eeq
where the free particle ground state $|\Omega\rangle$ is annihilated by all $a$ and $A$ operators.  To avoid clutter in the formulas, we will not normalize the states or probabilities.

These states are eigenstates of the bare Hamiltonian $H_0$.  We will need the following
\bea
H_0|L,p_2;H,p_1\rangle&=&\E_0(p_1,p_2)|L,p_2;H,p_1\rangle\nonumber\\
H_0|L,p_2;L,p_1;i,q\rangle&=&\E_{1i}(p_1,p_2,q)|L,p_2;L,p_1;i,q\rangle\nonumber\\
H_0|H,p_2;H,p_1;i,q\rangle&=&\tilde{\mathcal{E}}_{1i}(p_1,p_2,q)|H,p_2;H,p_1;i,q\rangle\nonumber\\
H_0|H,p_2;L,p_1\rangle&=&\E_2(p_1,p_2)|H,p_2;L,p_1\rangle \label{h0}
\eea
where the eigenvalues are
\bea
\E_0(p_1,p_2)&=&E_{SH}(p_1)+E_{DL}(p_2)\hsp
\E_{1i}(p_1,p_2,q)=E_{SL}(p_1)+E_{DL}(p_2)+e_{i}(q)\nonumber\\
\tilde{\mathcal{E}}_{1i}(p_1,p_2,q)&=&E_{SH}(p_1)+E_{DH}(p_2)+e_{i}(q)\hsp
\E_2(p_1,p_2)=E_{SL}(p_1)+E_{DH}(p_2). \label{e}
\eea

We will also need the action of the interaction Hamiltonian $H_I$ on some of the states
\bea
&&H_I|L,p_2;H,p_1\rangle\supset\sum_{i=1}^2 \int \frac{dq}{2\pi}\left(
\frac{|L,p_2;L,p_1-q;i,q\rangle}{\sqrt{8E_{SL}(p_1-q)E_{SH}(p_1)e_i(q)}}+
\frac{|H,p_2-q;H,p_1;i,q\rangle}{\sqrt{8E_{DL}(p_2)E_{DH}(p_2+q)e_i(q)}}
\right)\nonumber\\
&&\mathcal{P}H_I|L,p_2;L,p_1-q;i,q\rangle=\frac{|H,p_2+q;L,p_1-q\rangle}{\sqrt{8E_{DL}(p_2)E_{DH}(p_2+q)e_i(q)}}
\nonumber\\
&&\mathcal{P}H_I|H,p_2-q;H,p_1;i,q\rangle=
\frac{|H,p_2+q;L,p_1-q\rangle}{\sqrt{8E_{SL}(p_1-q)E_{SH}(p_1)e_i(q)}} \label{hip}
\eea
where $\mathcal{P}$ projects onto the subspace of the Fock space with one light source particle, one heavy detector particle and no other particles.  In applications to neutrino physics terms with other particles could correspond for example to processes in which virtual heavy particles are produced along with the neutrino.  At current reactor and accelerator experiments, such processes are strongly suppressed and so we do not consider these terms here.   In the first equation, we only considered the terms in the Fock space with precisely two $\phi$ particles and a neutrino, although actually there will be two more terms in the Fock space with four $\phi$ particles and a neutrino corresponding to a process in which the initial source and detector particles are both spectators and a neutrino is created along with either two source or two detector particles.  Needless to say, such a process will necessarily involve states which are far off-shell and so will have negligible contributions to probabilities. 

\section{Finite Time Probability Densities}

In particle physics and in particular in neutrino physics, the most commonly computed quantity is the S-matrix.  This is the amplitude for a decoupled state in the asymptotic past to evolve to a given decoupled state in the asymptotic future.  Of course our Universe is not believed to have an asymptotic past and in practice one is always interested in experiments which occur during a fixed time window, but the S-matrix nonetheless provides an extraordinarily accurate approximation to the scattering amplitudes of interest in terrestrial experiments.  It has been questioned whether the same reliability may be expected for neutrino oscillation experiments.  Here the question is quite subtle as the entanglement of the particles involved in neutrino production and detection with the environment is expected to play an important role, potentially invaliditating the assumed decoupling at early and late times which is an essential ingredient in the definition of the S-matrix.  We will not attempt to answer this question here.  Instead, we will compute a quantity which does not require decoupling.

In all local quantum theories, including quantum field theory, one can define finite time probabilities as follows.  Every quantum theory comes with a Hilbert space of states $\mathcal{H}$.  In the Schrodinger picture, on each time slice $t$ a given system is in one state $|t\rangle\in\mathcal{H}$.  Lorentz-invariant quantum field theories come with an operator $H$ which is time-independent in the Schrodinger picture and relates the states at times $t_i$ and $t_f$ via
\beq
|t_f\rangle=e^{-i(t_f-t_i)H}|t_i\rangle.
\eeq
The amplitude $\mathcal{A}$ for the evolution from the state $|\psi_1\rangle$ and time $t_i$ to a state $|\psi_2\rangle$ at time $t_f$ is given by first evolving $|\psi_1\rangle$ to time $t_f$ and then calculating its inner product with $|\psi_2\rangle$
\beq
\mathcal{A}=\langle \psi_2|e^{-i(t_f-t_i)H}|\psi_1\rangle.
\eeq
The corresponding probability $P$ is the norm squared of the amplitude $|\mathcal{A}|^2$.  In general states have continous quantum numbers and so the basis of states is infinite, often uncountable, and so all probabilities are formally equal to zero.  In practice this problem can be overcome by placing the system in a box so that the basis is countable and then multiplying the probability by a suitable prefactor so that it becomes a probability density with respect to some quantum numbers.  The prefactor is fixed by demading that the probability density integrates to unity.  Then one takes the limit as the size of the box goes to infinity.  

What are these probabilities physically?   The state of the Universe at time $t_i$ is fixed to be $|\psi_1\rangle$.  This is an initial value problem.  There is no decoupling assumption, nor are the particles assumed to be on-shell.  Such assumptions are not necessary as no infinite time limit is taken\footnote{The on-shell assumption is necessary in the definition of the S-matrix because the in and out states are assumed to be eigenstates of the free Hamiltonian $H_0$, as in the decoupling limit the evolution is generated by $H_0$ and so infinite-time limits only exist for $H_0$ eigenstates.}.   One may note that if time is run backwards from $t_i$ then one will arrive at a very different configuration.   At a future time $t_f$ the system is measured to see if it is in state $|\psi_2\rangle$.  Again, as no infinite time limit is taken, $|\psi_2\rangle$ does not need to have any decoupling properties nor does it need to be on-shell.  $P$ is the probability that the result of the measurement is affirmative.

A state is on-shell if it is the eigenstate of the free Hamiltonian.  In particle physics, one often states that a given particle is or is not on-shell.  Technically, a given particle does not have a well-defined energy, because energy is the eigenvalue of the Hamiltonian and the Hamiltonian acts on the whole state, not on individual particles.  It is common in quantum field theory calculations to nonetheless define an energy for each particle via a standard construction in complex analysis.  With such a definition in hand, one may ask whether a given particle is on-shell.  In our treatment, we do not use this construction and so we do not define an energy for each particle.  Nonetheless, one may ask whether a given state is an eigenstate of our free Hamiltonian, and so whether a state is on-shell.  We will see that, as expected from intuition from the path integral formalism, off-shell contributions lead to highly oscillatory contributions to the probability amplitudes which, when appropriately integrated, largely cancel, leaving a dominant contribution from on-shell states.

In the sequel we will set $t_i$ to $0$ and will denote $t_f$ simply by $t$.

\section{Analytic Results} \label{ansez}

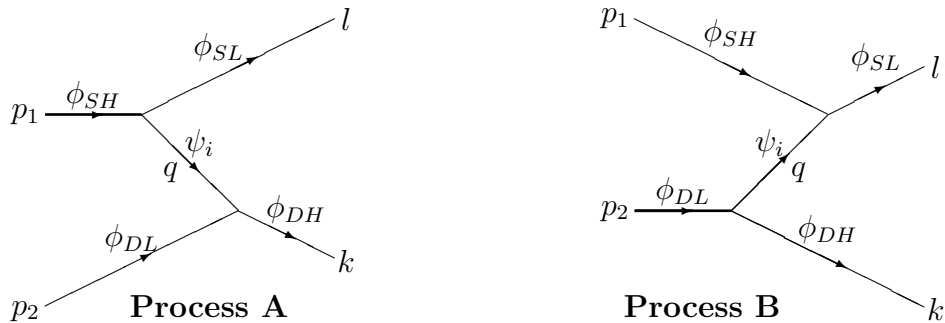
\begin{figure}
\setlength{\unitlength}{.5in}
\begin{picture}(3,4)(-4,-1)
\linethickness{1pt}
\put(-.3,-1){\makebox(0,0){{\bf{Process A}}}}
\put(-2,1){\vector(1,0){.6}}
\put(-1.5,1){\line(1,0){.5}}
\put(-1.5,1.2){\makebox(0,0){$\phi_{SH}$}}
\put(-2.2,1.0){\makebox(0,0){$p_1$}}
\put(-1,1){\vector(2,1){1.2}}
\put(0,1.5){\line(2,1){1}}
\put(-.2,1.7){\makebox(0,0){$\phi_{SL}$}}
\put(1.12,2){\makebox(0,0){$l$}}
\put(-1,1){\vector(1,-1){.6}}
\put(-.5,.5){\line(1,-1){.5}}
\put(-.4,.7){\makebox(0,0){$\psi_i$}}
\put(-.7,.4){\makebox(0,0){$q$}}
\put(0,0){\vector(2,-1){.6}}
\put(.5,-.25){\line(2,-1){.5}}
\put(.6,.01){\makebox(0,0){$\phi_{DH}$}}
\put(1.12,-.53){\makebox(0,0){$k$}}
\put(-2,-1){\vector(2,1){1.1}}
\put(-1,-.5){\line(2,1){1}}
\put(-1.1,-.29){\makebox(0,0){$\phi_{DL}$}}
\put(-2.2,-1.05){\makebox(0,0){$p_2$}}
\end{picture}
\begin{picture}(3,4)(-7,-1)
\linethickness{1pt}
\put(-1.3,-1){\makebox(0,0){{\bf{Process B}}}}
\put(-2,2){\vector(2,-1){1.2}}
\put(-1,1.5){\line(2,-1){1}}
\put(-1,1.85){\makebox(0,0){$\phi_{SH}$}}
\put(-2.2,2){\makebox(0,0){$p_1$}}
\put(0,1){\vector(2,1){.6}}
\put(.5,1.25){\line(2,1){.5}}
\put(.5,1.6){\makebox(0,0){$\phi_{SL}$}}
\put(1.12,1.5){\makebox(0,0){$l$}}
\put(-1,-.005){\vector(1,1){.6}}
\put(-1,-.005){\line(1,1){1}}
\put(-.6,.7){\makebox(0,0){$\psi_i$}}
\put(-.3,.4){\makebox(0,0){$q$}}
\put(-1,-.005){\vector(2,-1){1.2}}
\put(-1,-.005){\line(2,-1){2}}
\put(0,-.2){\makebox(0,0){$\phi_{DH}$}}
\put(1.12,-1.03){\makebox(0,0){$k$}}
\put(-2,-.005){\vector(1,0){.6}}
\put(-2,-.005){\line(1,0){1}}
\put(-1.5,.2){\makebox(0,0){$\phi_{DL}$}}
\put(-2.2,-.005){\makebox(0,0){$p_2$}}
\end{picture}
\caption{At leading order, contributions to the amplitude come from the same Feynman diagram which encodes two physical processes.  In process A (left) the source emits a neutrino which is absorbed by the detector.  In process B (right) a neutrino travels from the detector to the source.  The neutrino in process B is always off-shell.}
\label{procfig}
\end{figure}

We will consider the following experiment.  We begin with a heavy source particle and a light detector particle in initial wave packets\footnote{This should not be confused with the wave packet approximation for the neutrino that we avoid in our approach.  We believe that it is physically reasonable that the experimenter may prepare the source and detector wave packets with known properties at the beginning of the experiment.} at time $0$.  We measure the heavy detector particle at fixed time $t$.  In other words, at time $t$ our detector tells the experimenter whether the heavy detector particle is present.  We will work to leading nonvanishing order in $H_I$.  The leading order contribution comes at order $H_I^2$ at which two processes contribute, shown in Fig.~\ref{procfig}.  In process A, the heavy source particle decays to a light source particle and a neutrino.  The neutrino is then absorbed by the light detector particle, yielding the desired heavy detector particle.  In process B, the light detector particle emits a virtual neutrino and becomes a heavy detector particle.  The virtual neutrino is absorbed by the heavy source particle which becomes light.  In process B, the neutrino is necessarily well off-shell and so its contribution will always be small for a large separation between the source and the detector, but we keep it for completeness.

When is this leading order justified?  Clearly it is not possible to begin with $\phi_{SH}$ and $\phi_{DL}$ and finish with $\phi_{SL}$ and $\phi_{DH}$ with less than two interactions.  However it is possible with more.  Our approximation is justified when the two-interaction amplitude dominates over the sum of the multi-interaction amplitudes.   What could ruin such an approximation?  If the kinetic energy of the neutrino $E$ were much greater than the mass of the $\phi$ particles, then one would imagine that there would be many loop corrections of order $E^2/M_\phi^2$ which could be large.  We will always choose initial conditions that avoid this regime, since these corrections are not relevant at reactor neutrino experiments.  Furthermore if one considers a time interval such that the probability of $\phi_{SH}$ decay is of order unity, then higher order corrections reproduce the reduction in decay probability with time as the source is depleted\footnote{The reduction in the decay probability with source depletion arises in quantum field theory from loops in which $\phi_{SH}$ becomes $\phi_{SL}$ and $\psi$ and then again $\phi_{SH}$.  Each loop gives a factor of the self-energy $\Sigma$.  Summing the loops as a geometric series one adds a $\Sigma$ to the denominator of the $\phi_{SH}$ propagator.  The imaginary part of the new propagator yields the exponential decay at intermediate times, although a branch cut in the survival amplitude leads to a decay which is quadratic in time at very short times (the quantum Zeno effect) and polynomial at very long times \cite{tre,pasc1,pasc2}.}.
%\footnote{\bf "Source depletion" is also used to indicate the consumption of a macroscopic source. In those cases, usually, a classical approximation is sufficient to describe this effect: for example, in the case of a radioactive source, the total flux will decrease exponentially with time, to take into account the number of isotopes that have already decayed. Since in our case a single source particle is considered, the classical approximation cannot be used anymore and the decay probability (as a function of time) must be computed via Feynman diagrams. It is worth noticing that the two approaches, in the case of radioactive source, are equivalent (even if the microscopic treatment is formally more correct), indeed by resuming all the diagrams one obtains that the decay probability decrease exponentially. If only the leading order is considered (which is equivalent to say: the duration of the experiment t is shorter than the half-life of the source $\tau$), the probability decrease linearly with time, which is expected as long as $t\ll\tau$}
  Therefore the equations below cannot be trusted at time scales of order or greater than the half-life of $\phi_{SH}$.  In the case of reactor experiments, this means that our approximations will not describe the reactor fuel evolution, but can safely be applied to analyze the data on any given day.  This is sufficient for our purposes, as we are interested in the decoherence involved in individual events which last a small fraction of a second.  One day is much longer than any relevant time scale in this problem.   Needless to say, loop corrections to electroweak processes are rarely relevant in neutrino physics, but the above caveat is important for interpreting the large $t$ semianalytic approximation of Subsec.~\ref{tinf}.

The relevant amplitude for this process is
\beq
\mathcal{A}(k,l,p_1,p_2)=\langle H,k;L,l|e^{-iH t}|L,p_2;H,p_1\rangle .
\eeq
As the final state in this leading order process always consists of a heavy detector particle and a light source particle, we may insert the projection operator $\mathcal{P}$ from Eq.~(\ref{hip})
\beq
\mathcal{A}_i(k,l,p_1,p_2)=\langle H,k;L,l|\mathcal{P}e^{-iH t}|L,p_2;H,p_1\rangle .
\eeq

We are thus interested in those terms in
\beq
\mathcal{P}e^{-iH t}|L,p_2;H,p_1\rangle=\sum_{n=0}^{\infty}\frac{(-it)^n}{n!}c_n\hsp
c_n=\mathcal{P} (H_0+H_I)^n |L,p_2;H,p_1\rangle \label{stato}
\eeq
which have precisely two powers of $H_I$.  These are the leading terms in $c_n$ in perturbation theory.  In other words, if we chose to multiply $H_I$ by a parameter $g$ and expand in $g$, then the leading contribution to the amplitude would be of order $g^2$ and would be given by these terms.  We will drop higher order terms and, since all amplitudes will be proportional to $g^2$, we will simply drop the $g^2$ to avoid clutter.

The leading order terms are
\beq
c_n=\sum_{j=0}^{n-2}\sum_{k=0}^{n-j-2}\mathcal{P}H_0^j H_I H_0^k H_I H_0^{n-j-k-2}|L,p_2;H,p_1\rangle.
\eeq
Substituting in Eqs.~(\ref{h0}), (\ref{e}) and (\ref{hi}) one finds
\bea
c_n&=&\sum_{j=0}^{n-2}\sum_{k=0}^{n-j-2} \E_0^{n-j-k-2}(p_1,p_2)\int\frac{dq}{2\pi} \E_2^j(p_1-q,p_2+q)\label{ceq}\\
&&\times
\sum_{i=1}^2\left(\frac{\mathcal{E}_{1i}^k(p_1-q,p_2,q)+\tilde{\mathcal{E}}_{1i}^k(p_1,p_2+q,-q)}{8e_i(q)\sqrt{E_{SL}(p_1-q)E_{SH}(p_1)E_{DL}(p_2)E_{DH}(p_2+q)}}\right)
|H,p_2+q;L,p_1-q\rangle.\nonumber
\eea

Using the standard series expansion of the exponential and geometric series one arrives at the identity
\beq
\sum_{n=2}^{\infty}\sum_{j=0}^{n-2}\sum_{k=0}^{n-j-2}\frac{(-it)^n}{n!}\mathcal{E}_0^{n-j-k-2}\mathcal{E}_2^j\mathcal{E}_1^k
=\frac{1}{\E_0-\E_1}\left(\frac{e^{-i\E_0 t}-e^{-i\E_2 t}}{\E_0-\E_2}-\frac{e^{-i\E_1 t}-e^{-i\E_2 t}}{\E_1-\E_2}
\right) . \label{sumid}
\eeq
Let us define the shorthand notation
\beq
F(\E_0,\E_1,\E_2)=\frac{1}{\E_0-\E_1}\left(\frac{e^{-i\E_0 t}-e^{-i\E_2 t}}{\E_0-\E_2}-\frac{e^{-i\E_1 t}-e^{-i\E_2 t}}{\E_1-\E_2}\right).
\eeq
As a consistency check, note that this may be rewritten as an integral of $e^{-iEt}$ over the times $t_1$ and $t_2$ of the neutrino emission and absorption
\beq
F(\E_0,\E_1,\E_2)=(-i)^2\int_0^t dt_2\int_0^{t_2}dt_1 e^{-it_1 \E_0}e^{-i(t_2-t_1)\E_1}e^{-i(t-t_2)\E_2}. \label{tint}
\eeq
In the wave packet approaches of Refs.~\cite{beuthe,giunti2012}, the neutino emission time $t_1$ is fixed to be close to $0$ by imposing that the unobserved source particles, like $\phi_{SL}$, are in localized wave packets.  In our approach, on the other hand, $t_1$ can assume any value in the range $[0,t_2]$.  This is not imposed by hand, but is a result of our calculation of the Schrodinger picture evolution.

Inserting (\ref{ceq}) into (\ref{stato}) and performing the $j$, $k$ and $n$ sums using the identity (\ref{sumid}) we find the evolved state at time $t$
\bea
\mathcal{P}e^{-iH t}|L,p_2;H,p_1\rangle&=&\sum_{i=1}^2\int\frac{dq}{2\pi}
\left(
F(\E_0(p_1,p_2),\E_{1i}(p_1-q,p_2,q),\E_2(p_1-q,p_2+q))\right.\nonumber\\
&&
+
\left.F(\E_0(p_1,p_2),\tilde{\mathcal{E}}_{1i}(p_1,p_2+q,-q),\E_2(p_1-q,p_2+q))
\right)\nonumber\\&&
\frac{|H,p_2+q;L,p_1-q\rangle}{8e_i(q)\sqrt{E_{SL}(p_1-q)E_{SH}(p_1)E_{DL}(p_2)E_{DH}(p_2+q)}}.
\eea
The matrix elements are therefore
\bea
\mathcal{A}(k,l,p_1,p_2)&=&\sum_{i=1}^2 \mathcal{A}_i(k,l,p_1,p_2)\nonumber\\
\mathcal{A}_i(k,l,p_1,p_2)&=&\int\frac{dq}{2\pi}
\left(
F(\E_0(l+q,k-q),\E_{1i}(l,k-q,q),\E_2(l,k))\right.\nonumber\\
&&+\left. F(\E_0(l+q,k-q),\tilde{\mathcal{E}}_{1i}(l+q,k,q),\E_2(l,k))\right)\nonumber\\
&&\times\frac{\delta(p_2+q-k)\delta(p_1-q-l)}{8e_i(q)\sqrt{E_{SL}(l)E_{SH}(l+q)E_{DL}(k-q)E_{DH}(k)}}. \label{amp}
\eea

This is the amplitude for an initial momentum eigenstate, which is an infinitely extended plane wave.  To observe spatial neutrino oscillations, we must begin with localized states.  In particular, we would like the source and detector to be separated by a distance $x$.  Thus we will begin with the initial state
\beq
|0\rangle=\int dp_1 e^{-\frac{p_1^2}{2\sigma_1^2}}\int dp_2 e^{-\frac{p_2^2}{2\sigma_2^2}}e^{-ixp_2}|L,p_2;H,p_1\rangle
\eeq
where $\sigma_1$ and $\sigma_2$ are the uncertainties in the momenta of the initial source and detector particles.  These are in principle controlled and known by the experimenter, for example if the source is a radioactive particle in an optical trap.

%inverse size of the source and detector respectively.   Note that the size of the source and detector are known to the experimenter, and are necessary to understand the experiment.  For example, if the source size is of order an oscillation length, then the resulting oscillations will be smeared.  This effect is well-known and indeed reduces the sensitivity of JUNO to neutrino mass hierarchy by about 1$\sigma$ \cite{noijuno1,noijuno2}.

Therefore our initial states are described by wave packets.  Does this mean that we are using a wave packet model in the sense of Refs.~\cite{beuthe,giunti2012}?  Our claim is that we are not.  In those papers, the neutrino itself was described by a localized wave packet.  This is because it was assumed that $\phi_{SL}$ was measured, and the overlap between the trajectories of $\phi_{SL}$ and $\phi_{SH}$ was used to fix the neutrino production region.  Thus the neutrino was produced during a fixed, microscopic time.  On the other hand, in our approach $t_1$ is not fixed, rather it is integrated over a macroscopic time interval because our $\phi_{SL}$ is not measured.   As the neutrino production time is not fixed to a small interval, the neutrino wave function is not localized.   Therefore while our initial conditions for the source and detector are spatially-localized wave packets, our neutrino is not described by a spatially-localized wave packet and so our model is quite different from the wave packet models of Refs.~\cite{beuthe,giunti2012}.  We believe that our choice, in which the initial conditions are fixed and $\phi_{SL}$ is not measured, is appropriate for most if not all neutrino oscillation experiments today.

The evolution of this state is characterized by the amplitude
\bea
\mathcal{A}(k,l)&=&\langle H,k;L,l|e^{-iH t}|0\rangle=\int dp_1 e^{-\frac{p_1^2}{2\sigma_1^2}}\int dp_2 e^{-\frac{p_2^2}{2\sigma_2^2}}e^{-ixp_2}\mathcal{A}(k,l,p_1,p_2)\nonumber\\
&=&\sum_{i=1}^2\int\frac{dq}{2\pi}
\left(
F(\E_0(l+q,k-q),\E_{1i}(l,k-q,q),\E_2(l,k))\right.\nonumber\\
&&+\left. F(\E_0(l+q,k-q),\tilde{\mathcal{E}}_{1i}(l+q,k,q),\E_2(l,k))\right)\nonumber\\
&&\times\frac{e^{-\frac{(l+q)^2}{2\sigma_1^2}}e^{-\frac{(k-q)^2}{2\sigma_2^2}}e^{-ix(k-q)}}{8e_i(q)\sqrt{E_{SL}(l)E_{SH}(l+q)E_{DL}(k-q)E_{DH}(k)}}. \label{akleq}
\eea
The corresponding unnormalized probability density for a final state consisting of a heavy detector particle of momentum $k$ and a light source particle of momentum $l$ is
\beq
P(k,l)=|\mathcal{A}(k,l)|^2.
\eeq
The source particle is not observed, and so we are instead interested in the probability density to observe a detector particle with momentum $k$
\beq
P(k)=\int dlP(k,l). \label{pfin}
\eeq
Note that one may also define an appearance channel probability by multiplying the amplitude $\mathcal{A}_2$ in Eq.~(\ref{amp}) by $-1$, which would correspond to a sign flip in the $\phi_{DH}\phi_{DL}\psi_2$ interaction in $H_I$.  These unnormalized probability densities are proportional to the probability that the neutrino is detected at any time $t_2<t$, and so in this sense it is proportional to the time-integrated neutrino signal.  

\section{Numerical Results} \label{numsez}

\subsection{Fixing Parameters}

Imagine that the initial configuration is on-shell, for example with $p_1=p_2=0$.   As distance scales are macroscopic at neutrino experiments, we are interested in regimes where also the final state and even the intermediate neutrino are also on-shell.  This yields two conditions.  In the expression (\ref{amp}) for the amplitude, the neutrino momentum $q$ is integrated over.  Thus we effectively have one free parameter which needs to satisfy two conditions.  There is no solution for a generic choice of masses.    

This is one reason that we begin with the initial state $|0\rangle$.  Now $p_1$ and $p_2$ in principle run over all values, including those for which all particles are on-shell.  However, $\sigma_1$ and $\sigma_2$ combine to form the energy resolution of our measurement of neutrino oscillations and so if we wish to observe oscillations these should be small.  Therefore in general the on-shell contribution to the amplitude will be exponentially suppressed, unless the central value $p_1=p_2=0$ of the initial momentum distributions is close to a value which allows both the neutrino and the final states to be on-shell.

To remove this exponential suppression, we will choose the mass $M_{DL}$ so that for $p_1=p_2=0$ and $m_i=0$ all particles will be nearly on-shell for some value of $q$.  Such a tuning of $M_{DL}$ is very physical, it means that the detector is chosen to have an energy splitting which is close to the expected neutrino energy minus the recoil energy, which maximizes the probability that the neutrino will be detected.  Note that $M_{DL}=M_{HL}$ does not satisfy this condition as a result of the recoil energy.

Let us fix
\beq
p_1=p_2=0\hsp
M_{DH}=M_{SH}\hsp M_{SL}=M_{DH}(1-\epsilon)\hsp m_i=0. \label{masseq}
\eeq
As process B can never be on-shell, we only consider process A.

At the first vertex, a heavy source particle at rest decays to a light source particle and a neutrino with momentum $q$.  If all particles are on-shell, the initial and final energies at this vertex are
\beq
\E^1_i=M_{SH}\hsp
\E^1_f=q+\sqrt{M_{SH}^2(1-\epsilon)^2+q^2}.
\eeq
Energy conservation yields 
\beq
q\sim M_{SH}\left(\epsilon-\frac{\epsilon^2}{2}\right) \label{qsol}
\eeq
plus terms of order $\epsilon^3$.  Recall that $q$ is integrated over, and so the on-shell condition for these particles can always be satisfied for some $q$ in the range of integration.

At the second vertex, the neutrino is absorbed by a light detector particle at rest, which becomes a heavy detector particle.  If all particles are on-shell, the initial and final energies at this vertex are
\beq
\E^2_i=M_{DL}+q\hsp
\E^2_f=\sqrt{M_{SH}^2+q^2}.
\eeq
Energy conservation together with (\ref{qsol}) then yields
\beq
M_{DL}=M_{SH}\left(1-\epsilon+\epsilon^2\right). \label{mdl}
\eeq
As we would like a large on-shell contribution to our probabilities, we choose this mass for the light detector particle.  Equivalently, the detector mass splitting is chosen to optimize the chance that it absorbs a neutrino.

\subsection{Numerical Results}

We choose source and detector masses as in (\ref{masseq}) and (\ref{mdl}) with
\beq
M_{SH}=10\hsp
\epsilon=0.1.
\eeq
We also choose neutrino masses
\beq
m_1=0\hsp m_2=0.1
\eeq
and initial wave packet momentum spreads
\beq
\sigma_1=\sigma_2=0.1.
\eeq
We vary the separation $x$ between the source and the detector.  The detector measures the detector heavy particle momentum $k$, and so we are free to decide which value of $k$ interests us.  We fix
\beq
k=1
\eeq
as this is close to the on-shell value.  With these choices, the standard neutrino oscillation formula suggests oscillations with a wavelength of 
\beq
\lambda=4\pi \frac{|q|}{M_1^2-M_2^2}\sim400\pi. \label{lambda}
\eeq

\begin{figure} %[!tph]
\begin{center}
\includegraphics[width=2.5in,height=1.7in]{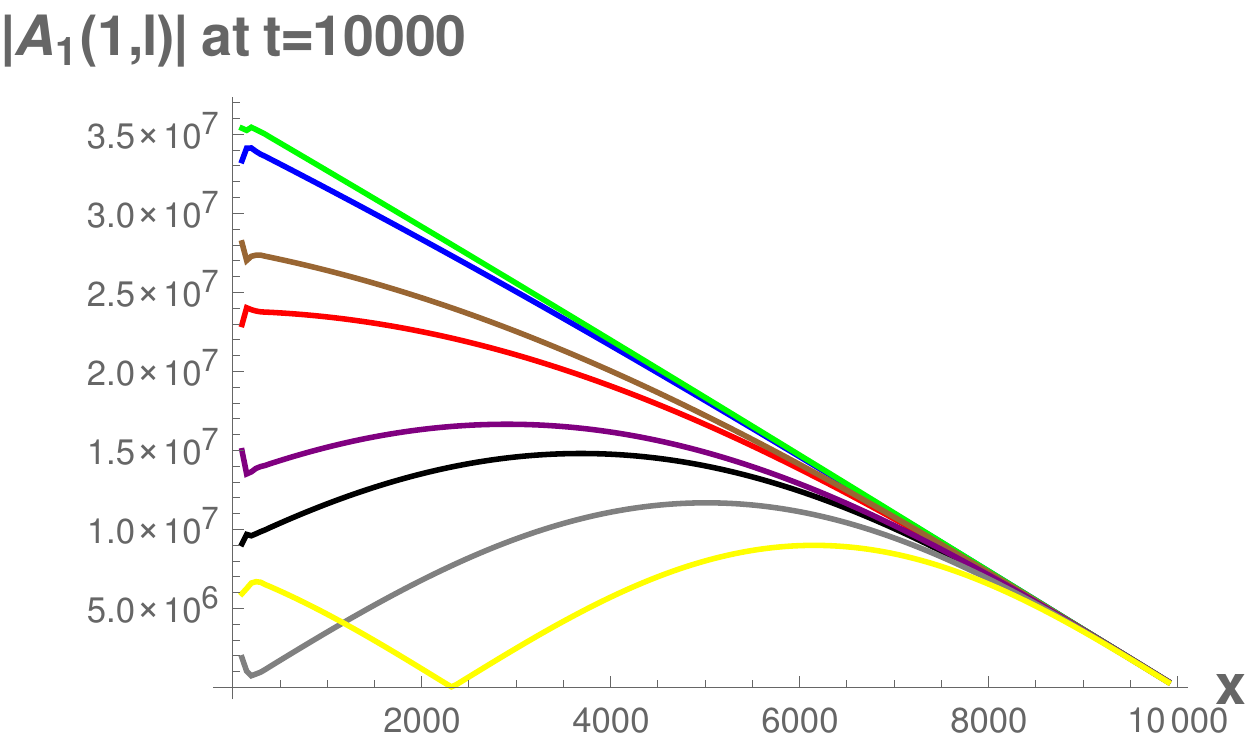}
\includegraphics[width=2.5in,height=1.7in]{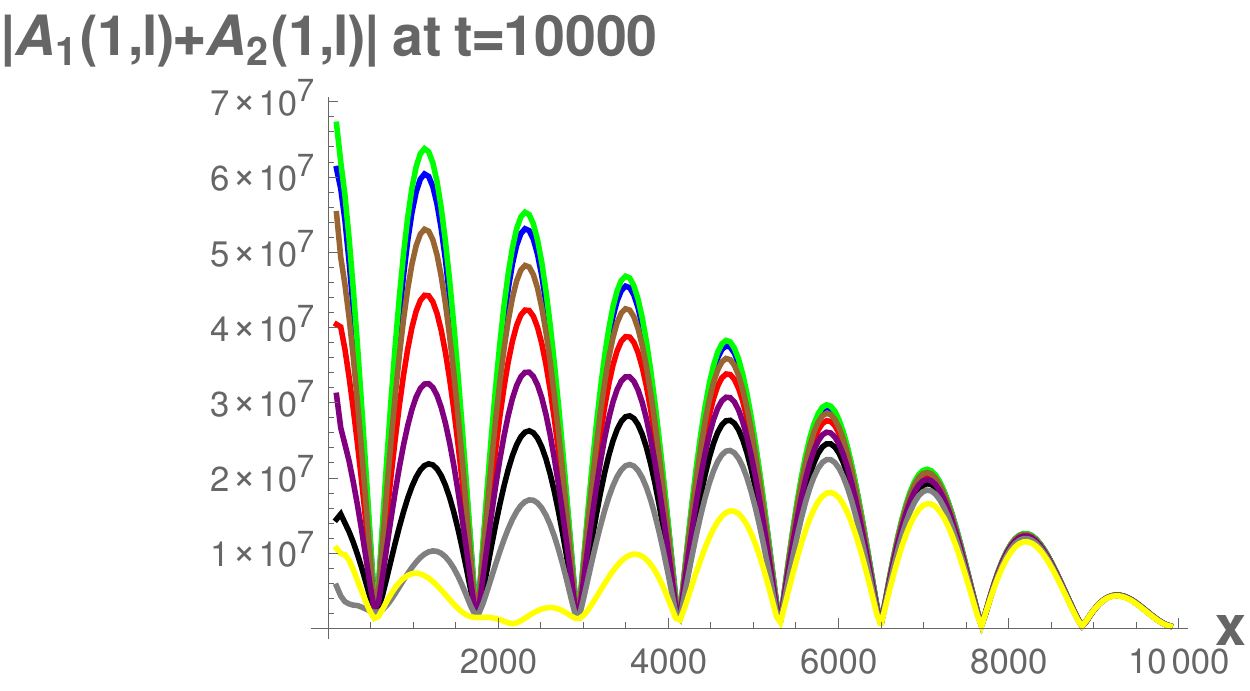}
\caption{Left: The one-flavor disappearance channel amplitudes $|A_1(1,l)|$ at $l=-0.942$ to $-0.956$ in steps of $-0.002$ in black, red, blue, green, brown, purple, grey and yellow respectively at time $t=10^4$.  Right: The sum $|A_1(1,l)+A_2(1,l)|$, which exhibits two-flavor oscillations.  In both cases, the highest curves are closest to the on-shell values.}
\label{afig}
\end{center}
\end{figure}

\begin{figure} %[!tph]
\begin{center}
\includegraphics[width=2.5in,height=1.7in]{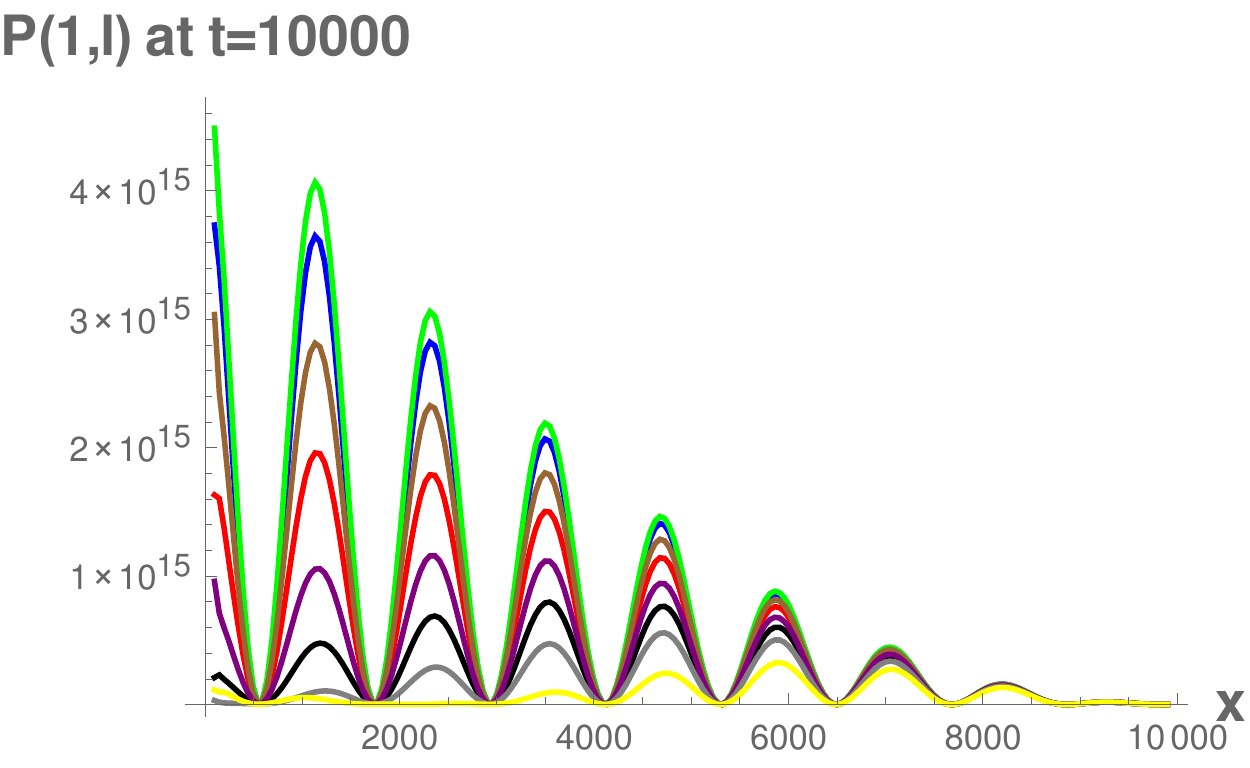}
\includegraphics[width=2.5in,height=1.7in]{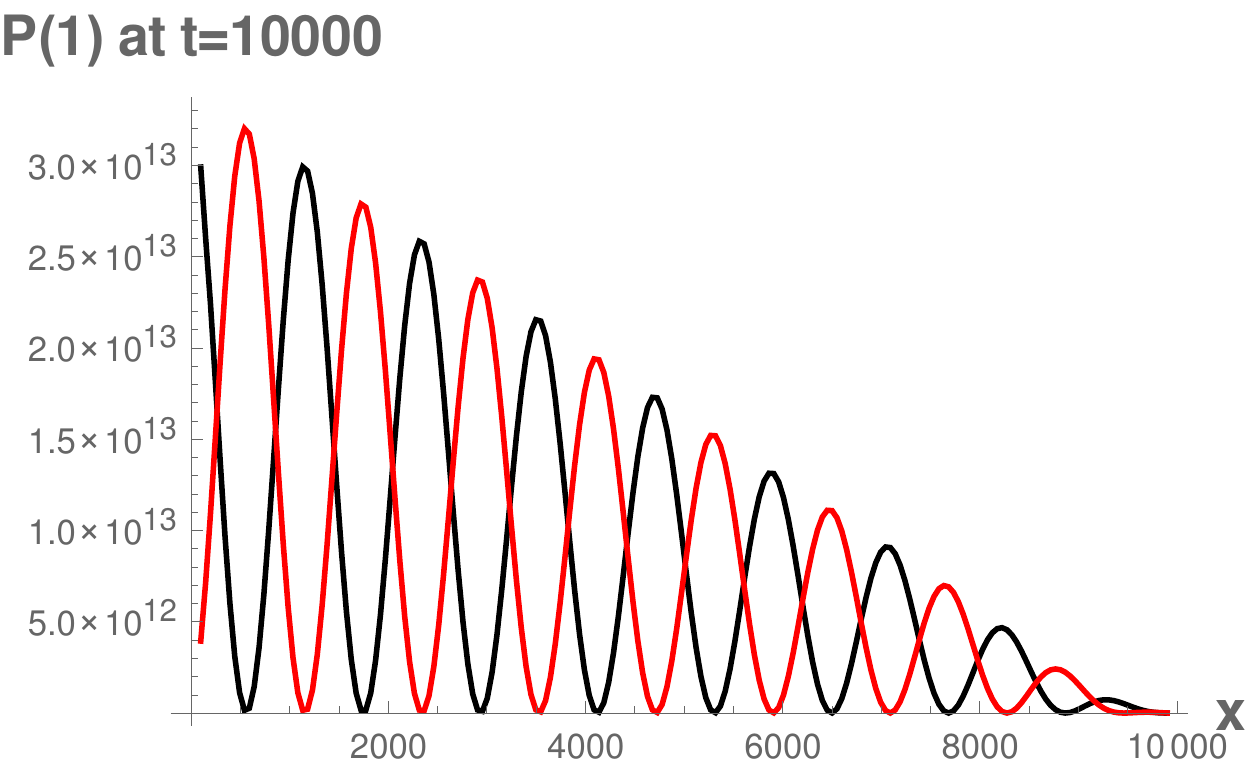}
\caption{Left: The disappearance channel probability density $P(1,l)$ at $l=-0.942$ to $-0.956$ in steps of $-0.002$ in black, red, blue, green, brown, purple, grey and yellow respectively at $t=10^4$.  Right: The disappearance (black) and appearance (red) oscillation probabilities $P(1)$ at time $t=10^4$.}
\label{pfig}
\end{center}
\end{figure}

The oscillation amplitudes and probabilities at time $t=10^4$ and varying baseline $x$ are shown in Figs.~\ref{afig} and \ref{pfig} respectively.  As can be seen on the left hand side of Fig.~\ref{afig}, near the on-shell value of $l$ the absolute value of the one flavor amplitude  shrinks linearly with $(t-x)$.  As a result, the two flavor amplitude exhibits oscillations whose amplitude shrinks linearly with $(t-x).$  This is because the neutrino production begins at $t=0$ and so neutrinos arrive at a distance $x$ between $t=x$ and $t=10^4$.  This time window shrinks linearly with $x$.   Thus the unnormalized probability $P(k)$, which is essentially the integrated neutrino flux, also shrinks linearly with $x$.  On the other hand, in the wave packet approximation one might expect instead a single bump at a value of $x$ equal to the time since production times the velocity.  The fact that no bump exists demonstrates that our production is not localized in time, as it would be in the wave packet approximation, but rather occurs continuously from time $0$ to $t$ as can be seen in Eq.~(\ref{tint}).  The oscillation wavelength is compatible with the standard formula (\ref{lambda}).  

\section{Semianalytic Estimates} \label{semisez}

\subsection{Infinite Time} \label{tinf}

The amplitudes simplify considerably if the time $t$ is infinite.   Of course our leading tree-level calculation is not valid if $t$ is of order or greater than the half life of $\phi_{SH}$, and so we will arrive at some nonsensical answers in this subsection such as infinite decay probabilities.  These will be remedied in the following subsection when we consider finite time.

We will now show that in this case the $t_1$ and $t_2$ integrations from Eq.~(\ref{tint}) each yield delta functions for momenta imposing that the interactions preserve the on-shell energy, which combined with the two delta functions from Eq.~(\ref{amp}) which imposed momentum conservation at the two vertices yield a total of four delta functions for the four momenta $p_1$, $p_2$, $q$, $k$ and $l$.  This means that only one momentum is independent and it allows all integrals to be trivially performed using the delta functions.

Let us begin with the $t\rightarrow\infty$ limit of Eq.~(\ref{tint}) 
\bea
F(\E_0,\E_1,\E_2)&=&-\lim_{t\rightarrow\infty}\int_0^t dt_2\int_0^{t_2}dt_1 e^{-it_1 \E_0}e^{-i(t_2-t_1)\E_1}e^{-i(t-t_2)\E_2}\nonumber\\
&=&-(2\pi)^2\lim_{t\rightarrow\infty}e^{-i t \E_2} \delta(\E_0-\E_1)\delta(\E_1-\E_2). \label{guscio}
\eea 
Recall that the functions $\E_i$ are the total energy that the system would have before, between or after the interactions were all particles on-shell, or equivalently $\E_i$ are the eigenvalues of $H_0$.  Of course the true energy, which is the eigenvalue of the total Hamiltonian $H$, is conserved for each $H$ eigenstate but these eigenstates are very complicated.  Therefore the fact that the $\E_i$ must be equal in Eq.~(\ref{guscio}) is a consequence of the fact that the states are on-shell in the infinite $t$ limit.  

Process B can never be on-shell, and so in the infinite time limit only process $A$ contributes to $\mathcal{A}$ in Eq.~(\ref{akleq})
\bea
\mathcal{A}(k,l)&=&-2\pi\lim_{t\rightarrow\infty}e^{-i t \E_2(l,k)}\sum_{i=1}^2\int dq 
\delta(\E_0(l+q,k-q)-\E_{1i}(l,k-q,q))\nonumber\\
&&\times\frac{\delta(\E_{1i}(l,k-q,q)-\E_2(l,k))e^{-\frac{(l+q)^2}{2\sigma_1^2}}e^{-\frac{(k-q)^2}{2\sigma_2^2}}e^{-ix(k-q)}}{8e_i(q)\sqrt{E_{SL}(l)E_{SH}(l+q)E_{DL}(k-q)E_{DH}(k)}}.
\eea
The two delta functions can be numerically solved to yield the values of $q$ and $l$ for which this process is on-shell for a given value of $k$.  In general there are two solutions, however with the choice of parameters in Sec.~\ref{numsez} one of these solutions is strongly suppressed by the exponential factors.   This suppression is caused by the small values of the momentum spreads $\sigma_i$ which are necessary for oscillations to be observed.    Let the remaining solutions be $q_i$ and $l_i$, which implicitly depend on $k$.  

In summary, the delta functions may be simplified by defining a function $f_i(k)$ such that
\beq
\delta(\E_0(l+q,k-q)-\E_{1i}(l,k-q,q))\delta(\E_{1i}(l,k-q,q)-\E_2(l,k))=f_i(k)\delta(q-q_i)\delta(l-l_i).
\eeq
The function $f_i(k)$ is given by the usual inverse Jacobian formula.  Then the $q$ integral may be performed, leaving  
\beq
\mathcal{A}(k,l)=-2\pi \lim_{t\rightarrow\infty}\sum_{i=1}^2 f_i(k)\delta(l-l_i)\frac{e^{-\frac{(l_i+q_i)^2}{2\sigma_1^2}}e^{-\frac{(k-q_i)^2}{2\sigma_2^2}}e^{-ix(k-q_i)}e^{-i t \E_2(l_i,k)}}{8e_i(q_i)\sqrt{E_{SL}(l_i)E_{SH}(l_i+q_i)E_{DL}(k-q_i)E_{DH}(k)}}.
\eeq
So far our amplitude is exact, at leading order in perturbation theory in the infinite time limit.  In practice, in neutrino oscillation experiments the various kinematic factors are, to within experimental error, equal for each relevant active neutrino mass eigenstate.  Therefore up to an over all constant, only the phase 
\beq
\mathcal{A}(k,l)\sim \sum_{i=1}^2 \delta(l-l_i) e^{ixq_i}e^{-i t \E_2(l_i,k)}
\eeq
is relevant for neutrino oscillations.   Here and from now on the infinite time limit will be implicit.
This yields an unnormalized probability density of
\beq
P(k,l)\sim  \left|\sum_{i=1}^2 \delta^2(l-l_i) e^{ixq_i}e^{-i t \E_2(l_i,k)}\right|^2=4\delta^2(l-l_s)\cos^2\left(\frac{x}{2}(q_2-q_1)-\frac{t}{2}(\E_2(l_2,k)-\E_2(l_1,k))\right).
\eeq
The integration over the unobserved momentum $l$ in Eq.~(\ref{pfin}) uses one delta function but leaves one $\delta(0)$ using the approximation that $l_i$ is independent of $i$.  This infinity is to be expected, as this probability density is proportional to the number of neutrinos measured after an infinite time with a constant neutrino flux, which of course is infinite.   Thus the observed neutrino flux at momentum $k$ will be proportional to
\beq
P(k)\sim 4\delta(0)\cos^2\left(\frac{x}{2}(q_2-q_1)-\frac{t}{2}(\E_2(l_2,k)-\E_2(l_1,k))\right). \label{inf}
\eeq
The time $t$ is infinite and so the phase has an infinite shift, but at fixed time nonetheless the $x$ wavelength is well-defined, it is
\beq
\Delta x=\frac{2\pi}{q_2-q_1}.
\eeq

In the case of the parameters of Sec.~\ref{numsez} the on-shell conditions are easily solved numerically to yield
\beq
q_1\sim 0.9497\hsp q_2\sim 0.9444\hsp \Delta x\sim 1183.
\eeq
This wavelength is again consistent with Fig.~\ref{pfig}.  Of course the overall shrinking of the oscillation amplitude with $x$ is not present in the infinite time case, as that effect resulted from the limited time for the neutrino beam to reach large $x$.  Instead the probability (\ref{inf}) is periodic.

\subsection{Finite Time}

In the finite time $t$ case, the on-shell condition is violated via a correction to the energy of order $1/t$.  This energy shift is much too small to be observed, but it is responsible for the space and time dependence of the unoscillated integrated flux.  The finite time case can be treated using a saddle-point approximation.  Eq.~(\ref{tint}) for $F$ is similar to the $e^{-iEt}$ used in the standard wave-packet approach \cite{beuthe,giunti2012} except that here the neutrino is created at $t_1$ and absorbed at $t_2$ which are integrated over all possible values between $0$ and $t$.  In the wave-packet approach, on the other hand, the creation time and absorption time are each fixed to within a small range.  As the time $t$ will be taken to be large, the states are still nearly on-shell and so we may linearly expand the various energies
\beq
\E_0=\epsilon_{0}+v_{0}(q-q_i)\hsp
\E_{1i}=\epsilon_{1i}+v_{1i}(q-q_i)\hsp
\E_2=\epsilon_{2}+v_{2}(q-q_i)
\eeq
where $q_i$ again are the on-shell values of $q$.  As the source and detector particles are very heavy, we will ignore their velocities and so set
\beq
v_0=v_2=0.
\eeq
Including these velocities would be straightforward.  The important velocity is that of the neutrino
\beq
v_{1i}=\left.\frac{\partial \E_{1i}(l,k-q,q)}{\partial q}\right|_{q_i}\sim \frac{q_i}{\sqrt{m^2_i+q_i^2}}.
\eeq
%In the ultrarelativistic limit relevant for Sec.~\ref{numsez}, one may approximate
%\beq
%v_{1i}\sim 1.
%\eeq
Finally, we will use the approximation
\beq
\epsilon=\epsilon_0=\epsilon_{1i}=\epsilon_2
\eeq
which holds when all particles are nearly on-shell.

\begin{figure} %[!tph]
\begin{center}
\includegraphics[width=2.5in,height=1.7in]{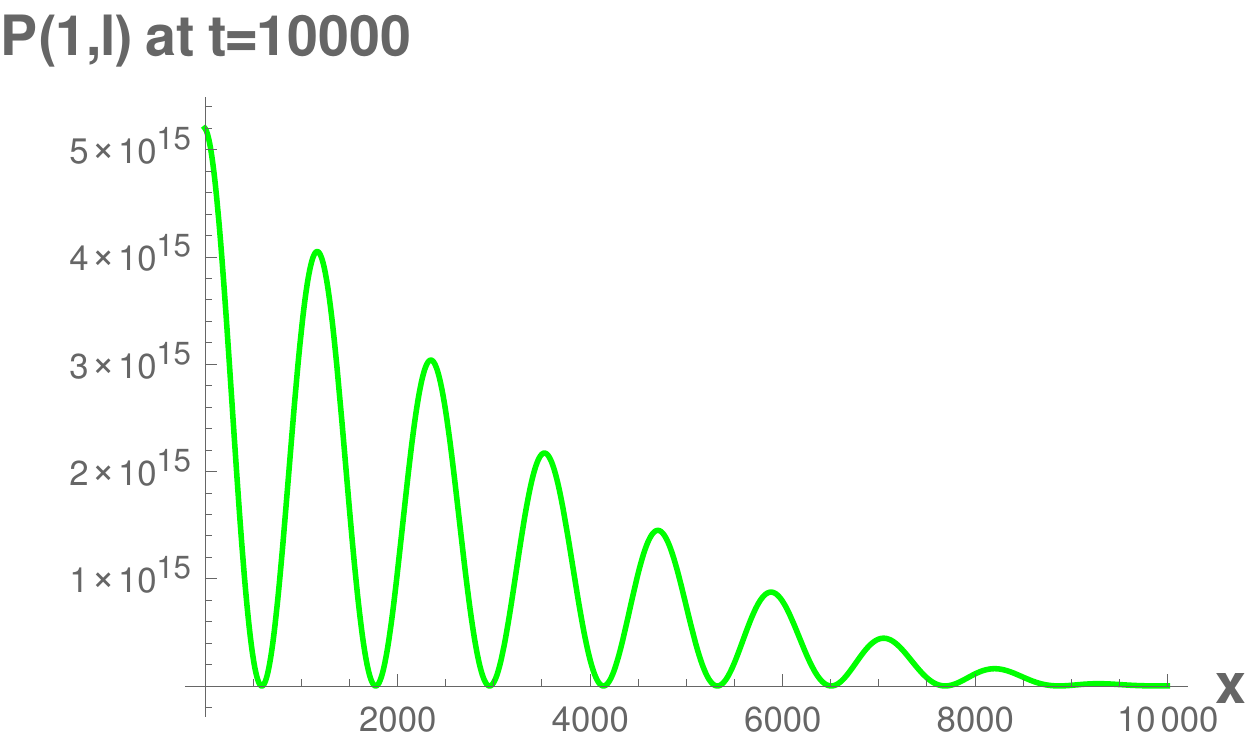}
\caption{The disappearance channel probability density $P(1,l)$ as approximated in Eq.~(\ref{papp}).}
\label{appfig}
\end{center}
\end{figure}

Again, only process A can be close enough to being on-shell to be relevant when $t$ is large.  $F$ in Eq.~(\ref{tint}) is then 
\beq
F(\E_0,\E_{1i},\E_2)\sim-e^{-i\epsilon t} \int_0^t dt_2\int_0^{t_2}dt_1 e^{-i(q-q_i)v_{1i}(t_2-t_1)}
\eeq
The $q$-dependent phases in Eq.~(\ref{akleq}) are then
\bea
\mathcal{A}(k,l)&\sim& \sum_i e^{iq_i x} \int_0^t dt_2\int_0^{t_2}dt_1\int dq e^{i(q-q_i)(x-v_{1i}(t_2-t_1))}\nonumber\\
&=& 2\pi\sum_i e^{iq_i x} \int_0^t dt_2\int_0^{t_2}dt_1 \delta(x-v_{1i}(t_2-t_1))\nonumber\\
&=& 2\pi\sum_i  \frac{e^{iq_i x}}{v_{1i}}\int_{x/v_{1i}}^t dt_2=2\pi\sum_i  \frac{e^{iq_i x}}{v_{1i}}\left(t-\frac{x}{v_{1i}}\right)\nonumber\\
&\sim&\sum_i e^{iq_i x}(t-x) \label{nol}
\eea
where the last line is the ultrarelativistic limit.  We have kept only the $q$-dependence of the phases and in particular we have dropped the $l$ dependence, although we kept the $l$ argument on the left-hand side of the equation for consistency of our notation.  Again summing and squaring one finds
\beq
P(k,l)\sim (t-x)^2\cos^2\left(\frac{x}{2}(q_2-q_1)\right). \label{papp}
\eeq
This is plotted in Fig.~\ref{appfig} with the $y$-axis rescaled to match the on-shell curves of the left panel of Fig.~\ref{pfig}.   Good agreement is found despite the crude approximations here, where our on-shell approximation effectively fixes $l$.   In the right panel of Fig.~\ref{pfig}, after $l$ has been integrated, the $(t-x)$ behavior is linear and not quadratic, as a result of the $l$-dependence of $P(k,l)$ which has been omitted in the approximation Eq.~(\ref{nol}), which kept only the $q$-dependent phases.  

More precisely, since the allowed distance of $l$ from the on-shell value of $l_i$ is inversely proportional to $(t-x)$, as is clear from the uncertainty principle, the $l_i$ integral is suppressed by a factor of $(t-x)$.  This suppression converts the quadratic $(t-x)$-dependence of $P(k,l)$ into the linear dependence of $P(k)$.   This is also evident in the left panel of Fig.~\ref{pfig}, as the fractional distance between the curves is largest on the left, where $(t-x)$ is largest, reflecting that the $l$ integration has narrower support at large $(t-x)$.

\section{Comparison with the Literature}  \label{litsez}

There is already a vast literature on neutrino decoherence.  In this section we try to place our work in the context of the wider literature.  The literature largely uses three approaches.  The first approach uses quantum mechanical models.  The second uses quantum field theory, but in a wave packet formulation.  The third uses quantum field theory directly, without imposing the existence of wave packets.

A general discussion and comparison of the first two approaches appears in Ref.~\cite{akqmqft}.   Here it is explained that wave packets emerge in quantum field theory via a restriction on the region of space time in which the neutrino was produced.  Neutrinos are produced in that paper via a 2-body decay, as in our work.  Therefore localized wave packets arise whenever the other particle produced during the decay, corresponding to $\phi_{SL}$ in our paper,  is detected.  The detection of this particle places it in a fixed wave packet given by the function $f_{P_f}$.  Such a detection of course never occurs for reactor neutrinos and is quite rare in general.  The case in which $\phi_{SL}$ is not measured is also briefly and qualitatively discussed in Subsec. 4.1 of Ref.~\cite{akqmqft}, however it is assumed that its momentum is known and so it is treated as a delocalized plane wave.  As it is delocalized, the neutrino production point is not known and so presumably wave packets  do not arise.  

This is similar to our approach, however we do not assume that the momentum of $\phi_{SL}$ is known precisely.  On the contrary, the momentum distribution of our $\phi_{SL}$ is simply determined by the evolution equations with fixed initial conditions for $\phi_{SH}$.  This results in entanglement between $\phi_{SL}$ and our neutrino, which we break by integrating the probability over all possible final states of $\phi_{SL}$ as these are not measured.  It is essential that we integrate the {\it{probability}} and not the amplitude because distinct $\phi_{SL}$ final states are distinct final states and thus must be summed incoherently.

The canonical reference on the second approach is Ref.~\cite{beuthe}, which contains a very general treatment, for example allowing the neutrinos to be unstable.   It is again assumed that the particles involved in the production are measured.  As a result, as in Ref.~\cite{akqmqft}, the {\it{amplitudes}} in their Eqs.~(27) and (28) are folded into a fixed wave function for these particles, called the overlap function.  The overlap function is completely arbitrary, but its size, together with the initial size of the source particle, fix the allowed space time region in which the neutrino is produced and so the size of the neutrino wave packet.  This of course is all in sharp contrast with our approach, in which only the initial condition is known and $\phi_{SL}$ is in no way measured or artificially restricted.  Therefore, although we do assign parameters $\sigma$ to our initial conditions, which we feel can be fixed and known by the experimenter, the allowed neutrino production region is macroscopic and so does not fit the general characterization of wave packet models in Ref.~\cite{akqmqft}.

Ref.~\cite{giunti2012} considers a special case, that relevant to neutrino oscillations, of that considered in Ref.~\cite{beuthe}.  The treatment is equivalent.  However there is one additional comment which gets to the heart of the issue.  The author writes that the overlap function for $\phi_{SL}$ may result from a measurement of $\phi_{SL}$ or else from   ``interactions ... with the surrounding medium.''  The first possibility is rarely or never realized in neutrino oscillation experiments.  And so the claim is that interactions with the medium are equivalent to a measurement of $\phi_{SL}$.  The goal of our program is precisely to test this assertion.

We are motivated in part by Refs.~\cite{dayadec,naumov1,naumov2}.   As these papers are written by overlapping authors and share essentially the same approach, let us discuss the first.  The amplitude (6) depends on the arbitrary parameters defined in Eq (5), which are not integrated over.  There is integration over the momenta of $\phi_{SL}$, but using the arbitrary weights from Eq (5) which are equivalent to fixing the outgoing state $\phi_{SL}$ as in the previous papers.  In Ref.~\cite{naumov2} the author arrives at a very different formula for wave packet spreading than that obtained in Ref.~\cite{beuthe}.  The difference is caused by a different choice of overlap functions, describing the state $\phi_{SL}$.  In our approach no such ambiguity occurs as we do not fix the wave function of $\phi_{SL}$.

The rest of our motivation comes from Ref.~\cite{revival}.  The third paragraph of Section II contains a very direct and interesting discussion of this issue: ``The interactions ... may be turned 'on' and 'off.' ... a real (microscopic) source or detector will be in an environment which is 'noisy,' so that the coherent emission or absorption of a neutrino gets cut off after some time due to the interactions of the source or detector with its surrounding environment." In other words, they restrict the neutrino production time by hand as a proxy for interactions with the environment.  This restriction leads to the revival of neutrino oscillations.  Our goal is to test whether or not such a cut is a reliable proxy for environmental interactions, by directly including environmental interactions and checking to see whether oscillations can be revived.

This revival claim was tested in quantum field theory in Ref.~\cite{giuntirevival}.  Here again it was assumed that all particles involved in production are measured.  This assumption is incorporated as usual by integrating the amplitude over a Gaussian kernel corresponding to the measured wave function for $\phi_{SL}$.

What about the third approach?

The approach of Ref.~\cite{cgl} is similar to ours.  They aim to construct a consistent treatment of neutrino oscillations in quantum field theory.  They explain that spin and three-body kinematics are not essential to understand oscillations, and so like us they consider a scalar model in which neutrinos are created in a two-body decay.  The full entanglement with the parent particles is kept but then is traced out to calculate probabilities, which is equivalent to the integration of probabilities in our paper.  They use the stationary phase approximation to extract the scaling of their probabilities.  This is sufficient for their goal of trying to derive the standard neutrino oscillation formula.  However we believe that our direct numerical evaluation of probabilities is more robust, and thus better suited to our goal of understanding decoherence.

Ref.~\cite{mcgreevy} also contains a consistent approach to neutrino oscillations in quantum field theory.  It uses observations only of quantities which are observed.  The authors in particular do not restrict the space time region in which the neutrino is created.  They conclude that, on the contrary to the usual wave packet description in which neutrinos of various  mass eigenstates are created simultaneously at roughly the same position and then separate, instead the heavier eigenstate is created before the light and they coalesce at the detector.  They conclude that the usual wave packet picture is wrong.  However environmental interactions are not included, and we suspect that, as argued in Ref.~\cite{revival}, these would spoil the coherence between neutrinos created at sufficiently distinct times and so affect their conclusions.  It is the goal of our project to test this suspicion.

In addition a number of papers have tried to estimate the wave packet size for various kinds of neutrino experiment.  The most common approach, used in Refs.~\cite{wilczek,rich,boriserr} is simply to assert that the time between interactions of the source is the coherence time and so the wave packet size should be this time multiplied by some velocity.  Different papers generally found very different wave packet sizes.  Needless to say, the time between interactions depends in general on the infrared cutoff, and without that cutoff is generally equal to zero as interactions occur continuously.  The first calculation of the wave packet size, in Ref.~\cite{nuss76}, included a phase shift from each interaction and assumed decoherence results when the shift is of order $2\pi$ radians.  The IR cutoff dependence thus disappears and this is well-defined.

All of these estimates are in contradiction with the observation in Ref.~\cite{noi1} that only the difference between interactions of the source before and after neutrino production causes decoherence.  For the solar neutrinos of interest in Ref.~\cite{nuss76} the difference is of order the interaction itself and so this is not a problem.  However, in the case of the $\beta$ decays relevant for reactor neutrinos, there is little difference between the parent and daughter source particle and so this difference may be one or two orders of magnitude smaller than the interaction itself.  Even more importantly, the arguments of Ref.~\cite{noi1} suggest that the interatomic forces which are usually used are too small to create noticeable decoherence at reactor experiments, but strong interactions inside of nucleii may be sufficient.  These would yield very different wave packet sizes from existing estimates, perhaps within the range of interest for JUNO.   It is the final goal of our line of research to find a sensible method for estimating the wave packet size.

\section{Conclusions}

In Ref.~\cite{noi1} we constructed a toy model of neutrino production, considering the full entanglement of the quantum states.  This entanglement exists in the full quantum field theory description, and so, as has been emphasized in Ref.~\cite{cgl}, should be considered in any calculation.  However the toy model of \cite{noi1} could not be connected, even qualitatively, to observations because it did not include neutrino measurement.  

In the current note that shortcoming has been remedied.  We have modified the model of \cite{noi1} to include neutrino detection and we have tested that it produces the standard cosine squared (\ref{papp}) behavior that is expected for maximal neutrino oscillations.  

There are many controversies in neutrino physics, for example regarding the observability of neutrino oscillation decoherence \cite{revival,mcdonald} and also whether neutrino mass eigenstates are created at the same point and then separate or are created at different points and then coalesce \cite{mcgreevy}, which are not specific to any model of electroweak interactions.  They depend only on the basic structure of the entanglement of the various wave packets.  However arguments in the literature often rely upon wave packet models, where these entanglements are not considered.  In particular interactions with the environment are generally included by simply projecting the state \cite{giunti2012}.  Our toy model manifests entanglement and yet is simple enough that the relevant probabilities can be expressed in closed form, albeit as integrals.  Entanglement with the environment can be added as in Ref.~\cite{noi1}.  Therefore we believe that our model is now ready to be applied to resolving these controversies, and more generally to understanding just when the wave packet approximation is and is not applicable.  

Is our probability (\ref{pfin}) equivalent to that obtained using quantum field theory with wave packets?  In that case, the neutrino flavor eigenstate is created in a localized space-time region, as a result of the fact that the particles involved in its production inhabit localized wave packets.  In the present case, we have seen in Eq.~(\ref{tint}) that our amplitude may be written as an integral over the production and detection times $t_1$ and $t_2$.  These integrals extend over the entire, macroscopic interval from time $0$ to $t$.  As a result the probability $P$, being the square of the amplitude, will contain cross-terms arising from distinct production times $t_1$.  In this sense, both the standard description in which the mass eigenstates are produced simultaneously and the paradigm of Ref.~\cite{mcgreevy} are included in our description.  This story will become more interesting in the sequel, when we couple to environmental interactions, as these will spoil the coherence in cross-terms with sufficiently different values of $t_1$.

\section* {Acknowledgement}

\noindent
JE is supported by the CAS Key Research Program of Frontier Sciences grant QYZDY-SSW-SLH006 and the NSFC MianShang grants 11875296 and 11675223.  EC is supported by NSFC Grant No. 11605247, and by the Chinese Academy of Sciences Presidents International Fellowship Initiative Grant No. 2015PM063.  JE and EC also thank the Recruitment Program of High-end Foreign Experts for support.

\end{document}

\bibitem{boya2011}
  D.~Boyanovsky,
  ``Short baseline neutrino oscillations: when entanglement suppresses coherence,''
  Phys.\ Rev.\ D {\bf 84} (2011) 065001
  doi:10.1103/PhysRevD.84.065001
  [arXiv:1106.6248 [hep-ph]].

\bibitem{accdec}
  B.~J.~P.~Jones,
  ``Dynamical pion collapse and the coherence of conventional neutrino beams,''
  Phys.\ Rev.\ D {\bf 91} (2015) no.5,  053002
  doi:10.1103/PhysRevD.91.053002
  [arXiv:1412.2264 [hep-ph]].

\bibitem{steven}
  Y.~L.~Chan, M.-C.~Chu, K.~M.~Tsui, C.~F.~Wong and J.~Xu,
  ``Wave-packet treatment of reactor neutrino oscillation experiments and its implications on determining the neutrino mass hierarchy,''
  Eur.\ Phys.\ J.\ C {\bf 76} (2016) no.6,  310
  doi:10.1140/epjc/s10052-016-4143-4
  [arXiv:1507.06421 [hep-ph]].

\bibitem{nuss76}
  S.~Nussinov,
  ``Solar Neutrinos and Neutrino Mixing,''
  Phys.\ Lett.\  {\bf 63B} (1976) 201.
  doi:10.1016/0370-2693(76)90648-1

\bibitem{wilczek}
  L.~Krauss and F.~Wilczek,
  ``Solar Neutrino Oscillations,''
  Phys.\ Rev.\ Lett.\  {\bf 55} (1985) 122.
  doi:10.1103/PhysRevLett.55.122

\bibitem{rich}
 J.~Rich,
  ``The Quantum mechanics of neutrino oscillations,''
  Phys.\ Rev.\ D {\bf 48} (1993) 4318.
  doi:10.1103/PhysRevD.48.4318

\bibitem{boriserr}
  B.~Kayser and J.~Kopp,
  ``Testing the wave packet approach to neutrino oscillations in future experiments,''
  arXiv:1005.4081 [hep-ph].

\bibitem{giunti2012}
 C.~Giunti,
  ``Neutrino wave packets in quantum field theory,''
  JHEP {\bf 0211} (2002) 017
  doi:10.1088/1126-6708/2002/11/017
  [hep-ph/0205014].

\bibitem{zurek}
  W.~H.~Zurek,
  ``Environment induced superselection rules,''
  Phys.\ Rev.\ D {\bf 26} (1982) 1862.
  doi:10.1103/PhysRevD.26.1862

\bibitem{cgl}
  A.~G.~Cohen, S.~L.~Glashow and Z.~Ligeti,
  ``Disentangling Neutrino Oscillations,''
  Phys.\ Lett.\ B {\bf 678} (2009) 191
  doi:10.1016/j.physletb.2009.06.020
  [arXiv:0810.4602 [hep-ph]].

\bibitem{akqft}
  E.~K.~Akhmedov and A.~Y.~Smirnov,
  ``Neutrino oscillations: Entanglement, energy-momentum conservation and QFT,''
  Found.\ Phys.\  {\bf 41} (2011) 1279
  doi:10.1007/s10701-011-9545-4
  [arXiv:1008.2077 [hep-ph]].

\bibitem{akmoss}
  E.~K.~Akhmedov, J.~Kopp and M.~Lindner,
  ``Oscillations of Mossbauer neutrinos,''
  JHEP {\bf 0805} (2008) 005
  doi:10.1088/1126-6708/2008/05/005
  [arXiv:0802.2513 [hep-ph]].

\bibitem{colemanlect}
 B.~G.~G.~Chen, D.~Derbes, D.~Griffiths, B.~Hill, R.~Sohn and Y.~S.~Ting,
  ``Lectures of Sidney Coleman on Quantum Field Theory,''
  doi:10.1142/9371 .

\bibitem{review}
  M.~Beuthe,
  ``Oscillations of neutrinos and mesons in quantum field theory,''
  Phys.\ Rept.\  {\bf 375} (2003) 105
  doi:10.1016/S0370-1573(02)00538-0
  [hep-ph/0109119].

\end{document}